\documentclass[preprint,12pt]{elsarticle}

\usepackage{epsfig}

\usepackage{latexsym,amsmath,amssymb,amstext}

\begin{document}

\title{Divergences in the quark number susceptibility : The origin and a cure}

\author[1]{Rajiv V. Gavai}
\author[2]{Sayantan Sharma\corref{cor1}}
\ead[2]{sayantans@quark.phy.bnl.gov}
\cortext[cor1]{Corresponding author, currently in Brookhaven National Laboratory}
\address[1]{Department of Theoretical Physics, Tata Institute of Fundamental
       Research, Homi Bhabha Road, Mumbai 400005, India.}
\address[2]{Fakult\"at f\"ur Physik, Universit\"at Bielefeld, 
           D-33615 Bielefeld, Germany}

\date{\today}
\begin{abstract}
Quark number susceptibility on the lattice, obtained by merely adding a $\mu N$
term with $\mu$ as the chemical potential and $N$ as the conserved quark number,
has a quadratic divergence in the cut-off $a$.  We show that such a divergence already 
exist for free fermions with a cut-off regulator.  While one can
eliminate it in the free lattice theory by suitably modifying the action, as is
popularly done, it can simply be subtracted off as well.  Computations of higher
order susceptibilities, needed for estimating the location of the QCD critical
point, then need a lot fewer number of quark propagators at any order.  We
show that this method of divergence removal works in the interacting theory.
\end{abstract}


\maketitle

\newpage

\section{Introduction}
The phase diagram of the strongly interacting matter described by Quantum
Chromodynamics(QCD) has been a subject of intense research in the recent years.
Usual weak coupling perturbative approach may work for sufficiently high
temperatures.  However, the gauge interactions are likely to be strong enough
for temperatures close to $\Lambda_{QCD}$, the typical scale of QCD,
necessitating strong coupling techniques.  Lattice QCD is the most successful
non-perturbative technique which has provided us with some interesting results
pertaining to the phase diagram. It is now fairly well known from independent
lattice studies that the transition from the hadron to the quark gluon plasma
phase at zero baryon density is a crossover ~\cite{tc1,tc2,tc3}.  At non-zero
density, or equivalently nonzero quark chemical potential $\mu$, one has to face
a sign problem : quark determinant is complex.  This does not allow for an
importance sampling based Monte Carlo study. Several ways have been advocated in
the recent years to circumvent the sign problem in QCD \cite{fodor1,imagmu,
cano1,taylor}.  From perturbative studies of model quantum field theories with the
same symmetries as QCD~\cite{pw} and chiral model investigations at $T<<\mu$~\cite{njl}, 
a critical end-point is expected in the QCD phase diagram.  If present, the critical-end point would
result in the divergence of the baryon number susceptibility.  Thus its Taylor
expansion~\cite{taylor} at finite baryon density as a series in $\mu_B/T$ can
be used to compute the radius of convergence, and therefore,  an estimate of the
location of the critical end-point~\cite{gg1,gg2}.  First such estimates of the
radius of convergence of the Taylor series have predicted the critical end-point
to be  at $T_E/T_c=0.94$ and $\mu_B/T_E=1.8(1)$ ~\cite{gg2}.  Recently, a study
on a finer lattice has suggested the continuum limit to be around
$T_E/T_c=0.94(1)~,~\mu_B/T_E=1.68(5)$~\cite{dgg}.  In the heavy-ion experiments,
the fluctuations of the net proton number could act as a proxy for the net
baryon number. The STAR experiment at Brookhaven National Laboratory has
reported the measurements for the fluctuations of the net proton number for a
wide range of center of mass energy $\sqrt s$, of the colliding heavy ions
between 7.7 and 200 GeV. At $\sqrt s=19.6$ GeV the experimental data are observed
\cite{star} to deviate from the predictions of the proton fluctuations for
models which do not have a critical end-point, and is similar to the lattice
QCD-based predictions \cite{ggplb} for a critical point, signaling its possible
presence. It would be thus important to have a thorough understanding of the 
systematics of the lattice QCD results and make them as much reliable as 
possible.

In addition to the usual suspects, such as continuum extrapolation or effects
due to the finiteness of the lattice spacing, the scale-setting, and the
statistical precision of the measurements, a key new important factor is that
the radius of convergence estimate requires ratios of as many higher orders of
quark number susceptibilities(QNSs) as possible. Currently the state of the art
is the eighth order QNS~\cite{gg1,gg2}. It is very important to verify whether
the existing results are stable if ratios of further higher order of QNS are
taken into account. In order to calculate the QNS of order $m$, one has to take
the $m^{th}$-derivative of the free energy with respect to the quark chemical
potential.  Since the popular method of incorporating the chemical potential on
the lattice is through $\exp(\pm \mu a)$ factors multiplying the forward and the
backward temporal gauge links respectively of the fermion
operator~\cite{hk,kogut}, there is an ever increasing proliferation of terms of
varying sign as $m$ increases.  Their large number as well as the large cancellations
amongst them at a specific order make it difficult to increase $m$ beyond eight
at present.  Introducing the chemical potential by a $\mu N$-term, where $N$ is
the corresponding conserved charge, leads to both much fewer terms and lesser
cancellations at the same $m$ \cite{gs}, thereby reducing the computational cost
up to 60 \% for terms up to the eighth order; more savings ought to accrue by
going to even higher orders.  Not only will this improve the precision of the
location of the critical point but more precise Taylor coefficients and more terms in the 
Taylor expansion can potentially also lead to a better control of the QCD equation of state 
at finite baryon density which will be needed for the analysis of the heavy-ion data from the 
beam energy scan at RHIC as well as the future experiments at FAIR and NICA.

In this paper, we discuss whether such a linear in $\mu$ approach is viable or
has unsurmountable problems by comparing with the usual exponential in $\mu$
method.  In section \ref{sec:freeth},  we revisit the number density for
non-interacting fermions in the continuum using a cut-off regulator.  We point
out that divergences appear already in the continuum free theory when the
cut-off regulator is taken to infinity contrary to the common knowledge.  We
then discuss an approach to tackle this divergence in the free theory.
By performing continuum extrapolation of
the second and fourth order QNS for quenched QCD for the linear approach,
we validate it in section \ref{sec:numres}.    This is the most important result of our paper.  
We discuss its possible consequences and the extensions to higher order QNS.

\section{Thermodynamics of non-interacting fermions}
\label{sec:freeth}
QCD thermodynamics can be derived from its partition function, written in the 
path integral formalism \cite{kapusta} as,
\begin{equation}
 \mathcal{Z}=\int \mathcal{D}A_\mu \mathcal{D}\bar \psi \mathcal{D} \psi
\rm{e}^{\int_0^{1/T} d\tau \int d^3 x\left[-1/2 Tr(F^2_{\mu,\nu})-\bar\psi
(\gamma_\mu(\partial_\mu-ig A_\mu)+m -\mu \gamma_4)\psi\right]},
\end{equation}
where $\psi$, $\bar \psi$ and $A_\mu$ represent the quark, anti-quark and 
the gluon fields respectively, whose the color indices are not written 
explicitly above.  $\mu$ is the chemical potential for the
the net quark number with the corresponding conserved charge being 
$\int d^3 x \bar \psi \gamma_4 \psi$.  Generalizations to various conserved
flavour numbers is straightforward.  For simplicity, we will consider only a
single flavour with the baryonic chemical potential $\mu_B = 3 \mu_q$.
Appropriate derivatives of $\mathcal{Z}$ lead to various thermodynamical
quantities, e.g., the quark number density, or equivalently (1/3) the baryon 
number density, is defined as,
\begin{equation}
 n= \frac{T}{V}\frac{\partial \ln \mathcal{Z}}{\partial \mu}|_{T=\text{fixed}}
\end{equation}
Earlier attempts to discretize the above theory to investigate the finite
baryon density physics on a space-time lattice revealed $\mu$-dependent
quadratic divergences in the number density and the energy density when the
chemical potential is introduced in the quark Dirac operator by multiplying it
with the corresponding conserved charge on the lattice.  These divergences,
which appear as a $\mu /a^2$ term in the expression for the
lattice number density with $a$ as the lattice spacing, are present even if the
gauge interactions are absent.  Through explicit calculation of the number
density for non-interacting fermions on the lattice, it was then
shown~\cite{hk,kogut,bilgav} that suitable modification of the $\mu N$ term in
the action, eliminates these divergent terms on the lattice, and yields a
finite $a \to 0$ continuum limit.  Numerical studies of the QNS for the
interacting theory subsequently confirmed that once the free theory divergences
are thus eliminated, no further divergences arise~\cite{gotl,ggquen}.  A
succinct way to describe all the various actions is to introduce functions
$f(\mu a) [g(\mu a)]$ as the multiplying factors for the forward (backward)
timelike gauge fields on the lattice.  While for the naive discretization, $ f
= 1 +a \mu$ and $ g =1-a \mu$ leads to a divergent baryonic susceptibility in
the continuum limit, the choice $f = \exp(a \mu)$ and $ g= \exp (- a \mu)$ does
not.

Clearly since all derivatives
of $f$ and $g$ are nonzero for the exponential case, whereas only the first
derivative is nonzero for the linear case, higher order QNS are a lot simpler
for the latter.  Furthermore, for fermions with better chiral properties such as
the Overlap fermions or the Domain Wall fermions, the exponential form leads to
a loss \cite{bgs} of the exact chiral symmetry on lattice for nonzero $\mu$.  Indeed the 
only chiral symmetry preserving form these fermions have for finite $\mu$ and $a$ is the 
linear form \cite{sharma}. This motivates us to revisit the
issue of the nature and origin of these divergences when the chemical potential
enters linearly instead of the exponential form.  As we show below, the
divergences are present for the continuum free fermions as well, and the lattice regulator
simply faithfully reproduces them. While one can employ the freedom of lattice
action to eliminate them, it is not necessary. Indeed, one can perhaps employ
simpler subtraction methods to eliminate them, as we demonstrate in this paper.

\subsection{Continuum free fermions}
\label{sec:freecn}
\begin{figure}
\begin{center}
\includegraphics[scale=0.5]{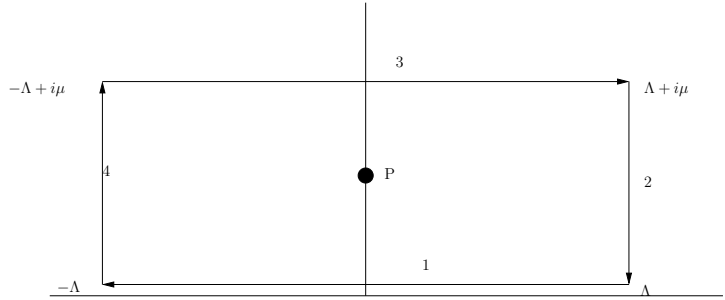}
\caption{The contour diagram for calculating the number density for free 
fermions at zero temperature. P denotes the pole. }
\label{contour}
\end{center}
\end{figure}

Results for the continuum free fermions are easily found in textbooks
\cite{kapusta}.  We review them below solely with the idea of pointing out
explicitly the $\mu$-dependent divergences present in them.  For simplicity, we
consider only massless fermions though this derivation can be easily extended
for finite mass.  The expression for the number density for free fermions is
easily obtained from the definitions above as 
\begin{equation}
 \label{eqn:nodensity}
 n=4iT\sum_{j=-\infty}^{\infty}\int
\frac{d^3p}{(2\pi)^3}\frac{(\omega_j+i\mu )}
 {p^2+(\omega_j+i\mu )^2}\equiv4iT\sum_{j=-\infty}^{\infty} F(\omega_j,\mu)~,
\end{equation}
where $p^2= p_1^2+ p_2^2+ p_3^2$ and $\omega_j=(2j+1)\pi T$. 
Here we choose the gamma matrices to be all Hermitian as is common in lattice
studies.  The continuum convention followed in the standard texts has only
$\gamma_4$ as Hermitian and the other gamma matrices are anti-Hermitian.
The expression in Eq.  (\ref{eqn:nodensity}) can be evaluated by the usual trick of converting the 
sum over energy states to a contour integral. The Matsubara frequencies lie on the real $\omega$-axis. 
Following \cite {kapusta} again, one can employ an infinitesimally small 
contour around the each pole on the real $\omega$ axis to represent the
$\omega_j$-sum, and obtain
\begin{equation}
\label{eqn:contlint}
2 \pi T\sum_jF(\omega_j,\mu)=Lt_{\epsilon\rightarrow0}\left[
\int_{-\infty+i\epsilon}^{\infty+i\epsilon} \frac{F(\omega,\mu) d\omega}{\rm{e}^{i\omega/T}+1}
+ \int_{\infty-i\epsilon}^{-\infty-i\epsilon} \frac{F(\omega,\mu) d\omega}{\rm{e}^{i\omega/T}+1}
\right]~.
\end{equation}
The line integrals in Eq.  (\ref{eqn:contlint}) can in turn be written in terms of contours in the upper 
and lower complex $\omega$ planes. Using the exact identity,
\begin{equation}
\label{eqn:imwsplt}
\frac{F(\omega,\mu)}{\rm{e}^{i\omega/T}+1}=F(\omega,\mu)-
\frac{F(\omega,\mu)}{\rm{e}^{-i\omega/T}+1}~,
\end{equation}
the line integral in the Im$\omega>0$ plane at infinity can be made convergent. The number density expression 
at finite temperature and density is,
\begin{equation}
\label{eqn:ncont0}
 n=\frac{2 i}{\pi}\left[\oint_{Im \omega<0}\frac{F(\omega,\mu)d \omega}{\rm{e}^{i\omega/T}+1} -
 \oint_{Im \omega>0}\frac{F(\omega,\mu)d\omega}{\rm{e}^{-i\omega/T}+1}
 +\int_{-\infty}^{\infty} F(\omega,\mu) d\omega\right]~.
\end{equation}
We note that the last term of the above expression contributes to number 
density at all temperatures, $T$ and $\mu$ and has only $\mu$-dependence. 
The first two terms yield the usual Fermi-Dirac distribution functions.
These have no ultraviolet divergences ~\cite{gpy} since the ultraviolet modes are exponentially 
suppressed. In order to examine the last term in detail, let us write it explicitly:
\begin{equation}
\label{eqn:ncont}
n= 4i \int_{-\infty}^{\infty} \frac{d \omega}{2\pi} 
\frac{d^3p}{(2\pi)^3}\frac{(\omega+ i\mu )} {p^2+(\omega+ i\mu )^2}~.
\end{equation}
Under a variable transformation $\omega+i \mu =\omega'$, it can be recast as
\begin{equation}
\label{eqn:ncont1}
n= 4i\int_{-\infty+i\mu}^{\infty+i\mu} \frac{d \omega'}{2\pi} 
\frac{d^3p}{(2\pi)^3}\frac{\omega'} {p^2+\omega'^2}~.
\end{equation}
In order to compute the integral carefully, we impose a cut-off $\Lambda$ on all
the four momenta and use the contour in the complex $\omega$ plane in
Figure \ref{contour}, leading to

\begin{equation}
\label{eqn:ncont2}
 n = 2i\int\frac{d^3p}{(2\pi)^3}\left[-i \Theta
\left(\mu- p \right) -\Bigg( \int_2+\int_4+\int_1 \Bigg) \frac{d \omega}{\pi} 
\frac{\omega} {p^2 +\omega^2} \right]~.~
\end{equation}

While the first term arises from
the residue of the pole of the integrand enclosed by the contour, the last three
terms in the Eq.(\ref{eqn:ncont2}) arise from closing the contour.  The line
integral 1 is identically zero because the integrand is an odd function.  The
sum of the line integrals 2 and 4 is
\begin{equation}
\label{eqn:ncont5}
 \int_{2+4}=-
 \frac{1}{2\pi} \ln \left[ \frac{p^2+(\Lambda+i \mu)^2}{p^2+(\Lambda-i
\mu)^2}\right].
\end{equation}
Since $\Lambda>>\mu$, $\Lambda$ being the cut-off, expanding the logarithm and 
retaining the leading term, 
\begin{equation}
\label{eqn:ncont6}
 \int_{2+4}= -\frac{4 i \mu \Lambda}{2 \pi (\Lambda^2+p^2)}~.
\end{equation} 
It is straightforward to do the remaining momentum integrals in
Eq.  (\ref{eqn:ncont2}).  The first term leads to the usual $\mu^3$ term.
However, the sum of the two line integrals 2 and 4 in Eq. (\ref{eqn:ncont6})
yields a $\mu \Lambda^2$ divergence in the expression for number density as below,
\begin{eqnarray}
 \nonumber
 -\int_0^\infty \frac{dp}{2 \pi^3} \frac{4 \mu  p^2 \Lambda}{\Lambda^2+ p^2}&=&
 -4 \mu \Lambda\int_0^\Lambda \frac{dp}{2 \pi^3}\left[ 1-\frac{\Lambda^2 }{\Lambda^2+ p^2}\right]\\\nonumber
&=&
-\frac{2\mu \Lambda^2}{\pi^3}\left[1-\frac{\pi}{4}\right]~.
\end{eqnarray}
Note that the leading diverging $\Lambda^3$-contribution, present for $\mu=0$, is the same for the line integrals 
2 and 4, and does cancel.   It is the non-leading $\mu$-dependent term which leads to the $\Lambda^2$ divergence above.  
This divergence shows up  if one uses Pauli-Villars method as well.  One then has 
 to introduce additional Pauli-Villars fields to cancel this $\mu\Lambda^2$ 
 divergence from the free energy distinct from those which are required to 
 cancel the usual $\Lambda^4$ divergence. 

Similarly, the lattice as a cut-off regulator for the free theory also leads
to $\mu a^{-2}$ divergence.
Using nonzero $T$ as the regulator, following the method outlined in~\cite{kapusta},
yields the contributions of only the two terms in Eq. (\ref{eqn:ncont0}).
This choice of the regulator does not permit the $T$-independent term of 
Eq. (\ref{eqn:ncont0}).  Such a choice
of regulator is, however, not feasible for lattice QCD computations, and indeed,
many other interacting theories.  
In order to obtain physically meaningful result, the contribution of the
line integrals 2 and 4 has to be subtracted off.  This free theory subtraction,
though $\mu$-dependent is analogous to the subtraction from the pressure at $T
\neq 0$, commonly used on the lattice in the equation of state computations.
We show in the  next section that no further divergences are observed
 once the free theory divergence is subtracted from the quark number susceptibilities.

\section{Quenched results on the lattice}

\label{sec:numres}
The analytical proof of  Refs. \cite{hk,kogut,gavai} for the lack of divergences
in the quark number susceptibility in the exponential $\mu$ case , outlined
briefly above in the previous section, was for non-interacting fermions.  No 
equivalent proof exists for the interacting fermions even for the exponential
case.   On the other hand, it is easy to check that for the staggered quarks, 
the chiral symmetry is maintained for $\mu \ne 0$.  For
the linear $\mu$ case, one even deals with conserved baryonic currents on the
lattice, $a \mu$ being the coefficient of the conserved baryon number on the
lattice.  One therefore expects no further divergences to arise, and no extra
renormalization needed, after switching on the gauge interactions in either
case.  This was explicitly checked by numerical simulations in the exponential
(indeed, generically for any $f \cdot g = 1 $) case for QCD in the quenched
approximation \cite{ggquen,swagato}.  It was proposed in Ref. \cite{gs,gs12},
and again demonstrated for the non-interacting fermions, that the spurious and
divergent terms arising in the linear $\mu$ case can be evaluated and explicitly
subtracted.  Clear tests of such a proposal for the interacting case are that
the continuum limit of $a \to 0$ of the so-subtracted quark number
susceptibilities should i) exhibit no additional divergences and ii) yield the
same result as for the exponential $\mu$ form.  In this section, we report
results of our these numerical tests for the linear $\mu$ case in quenched QCD
and verify that it passes both the tests, as expected.  The choice of the
quenched approximation was governed by the fact that the corresponding published
available results \cite{ggquen,swagato} for the exponential case make it simpler
to compare.  

On an $N^3 \times N_T$ lattice, the temperature is given by $T = 1/(N_Ta)$,
where $a$ the lattice spacing is governed by the gauge coupling $\beta = 6/g^2$.
We employ standard Wilson plaquette action for the gauge fields and use the
staggered quarks for our susceptibility determinations with the corresponding
Dirac matrix $D(a \mu)$ given by
\begin{eqnarray}
\nonumber
 &&D(\mu)_{xy}= \sum_{i=1}^3 \left[
\eta_i U_i(x)\delta_{x,y-\hat{i}}-\eta_iU^{\dagger}_i(y)\delta_{x,y+\hat{i}}\right]  \\
            &  - &
(1 -\mu a)\eta_4U^{\dagger}_4(y)\delta_{x,y+\hat{4}}+ (1 +\mu a)
\eta_4 U_4(x) \delta_{x,y-\hat{4}} + m a~ \delta_{x,y}.
\label{eqn:diracop}
\end{eqnarray}
Here $ma$ is the quark mass and $\eta$'s are the usual staggered fermion phases.
Our quenched QCD configurations were generated by using the Cabibbo-Marinari
pseudo-heatbath algorithm with three SU(2) subgroup update per sweep using the
Kennedy-Pendleton updating method.   We chose to simulate at two different
temperatures and two different quark masses on a variety of lattice sizes, as
listed in the Table \ref{tab:table1}, by selecting suitable $\beta$ values
\cite{ggquen,swagato} such that $T$ was held constant as $N_T$ (or equivalently
$a$) was increased (decreased).  This enabled us to make a continuum limit
extrapolation at both the temperatures.  We quote the temperatures in the units
of the critical temperature $T_c$ corresponding to the first order transition
for SU(3), defined by using the order parameter, the Polyakov loop.  Although we
do not need it explicitly anywhere below, we mention that $T_c= 276(2)~$
MeV~\cite{tcsu3} in the continuum limit, using the string tension value to be
425 MeV to set the scale.  Noting from Eq. (\ref{eqn:diracop}) that only the
first derivative with respect to $a \mu$, $D'$, is nonzero, the quark number
susceptibility for this linear chemical potential action is :
\begin{equation}
 \chi_{20}=\frac{T}{V}\left[\langle tr(-D^{-1}D'D^{-1}D')\rangle+
 \langle tr(D^{-1}D')^2\rangle \right]
\end{equation}
This expression is similar to that in Ref. \cite{gg1} but without the $D''$ term
which is identically zero here, as are further higher derivatives with $\mu$.
Our notation is same as in Ref. \cite{gg1} to facilitate comparison of our
numerical results below.  The traces in the above expression were computed
stochastically using Gaussian random vectors.  From the Monte Carlo time
evolution of the different operators that enter the susceptibility computation,
including the two terms above separately, we estimate that the autocorrelation
length is much less than $1000$ sweeps.  In order to ensure that our measurements
are statistically independent, they were done on configurations \footnote{The
configurations generated were rotated to the zero Polyakov loop sector to make
them similar to full QCD.} separated by $1000$ heatbath sweeps and excluding the
first $5000$-$10000$ sweeps for thermalization.   Such $N_{configs}$ configurations,
which varied from $24$-$100$, were employed to obtain the thermodynamic averages.
The details of the number of random vectors and number of configurations used 
at each quark mass and temperature are summarized in the Table \ref{tab:table1}.

As discussed in the previous section, the choice of $D(a \mu)$ in Eq.
(\ref{eqn:diracop}) leads to a QNS with a term $\propto1/a^2$ for the free case.
At each value of lattice cut-off, we computed numerically the coefficient for
this $1/a^2$-term for non-interacting fermions on the corresponding $N^3 \times
\infty$ lattice and subtracted it from the computed values of the susceptibility
in the interacting case.  If there are no additional divergences in the
interacting theory, one expects the continuum extrapolation in $1/N_T^2 \sim
a^2$ performed on these subtracted values of the susceptibilities to have a
smooth limit. On the other hand, if further divergences do exist in the
interacting theory then the $1/a^2$ or equivalently for a fixed temperature the
$N_T^2$ dependent term would survive and increase rapidly to blow up in the
continuum limit.

\begin{table}
 \centering
 \begin{tabular}{|c|c|r|r|r|r|r|}
 \hline \hline
 $T$ & $m/T_c$ & $\beta$ & $N$ & $N_T$ & $N_{configs}$ & $N_{rvec}$ \\
 \hline
 1.25 $T_c$& & 5.788 & 16 &4 & 100 &500\\
 & & 6.21 &24 &8 &50 & 500\\
 & 0.1 & 6.36 &30 &10 &60 &500\\
 & & 6.505 &36 &12 &24 &500\\
 \hline
 1.25 $T_c$& & 5.788 & 16 &4 & 100 &500\\
 &0.01 & 6.21 &24 &8 &48 & 500\\
 &  & 6.36 &30 &10 &68 &500\\
 &   &6.505&36 &12 & 24  &500\\
 \hline \hline
 & & 6.0609 & 16 &4 & 100 &400\\
  & & 6.3331 & 32 &6 & 50 &400\\
 2 $T_c$ & 0.1 & 6.45 &24 &8 &80 & 400\\
 &  & 6.75 &22 &10 &80 &400\\
 \hline
  \end{tabular}
  \caption{The parameters for the lattice simulations}
  \label{tab:table1}
\end{table}

\begin{figure}[h]
\begin{center}
\includegraphics[scale=0.52]{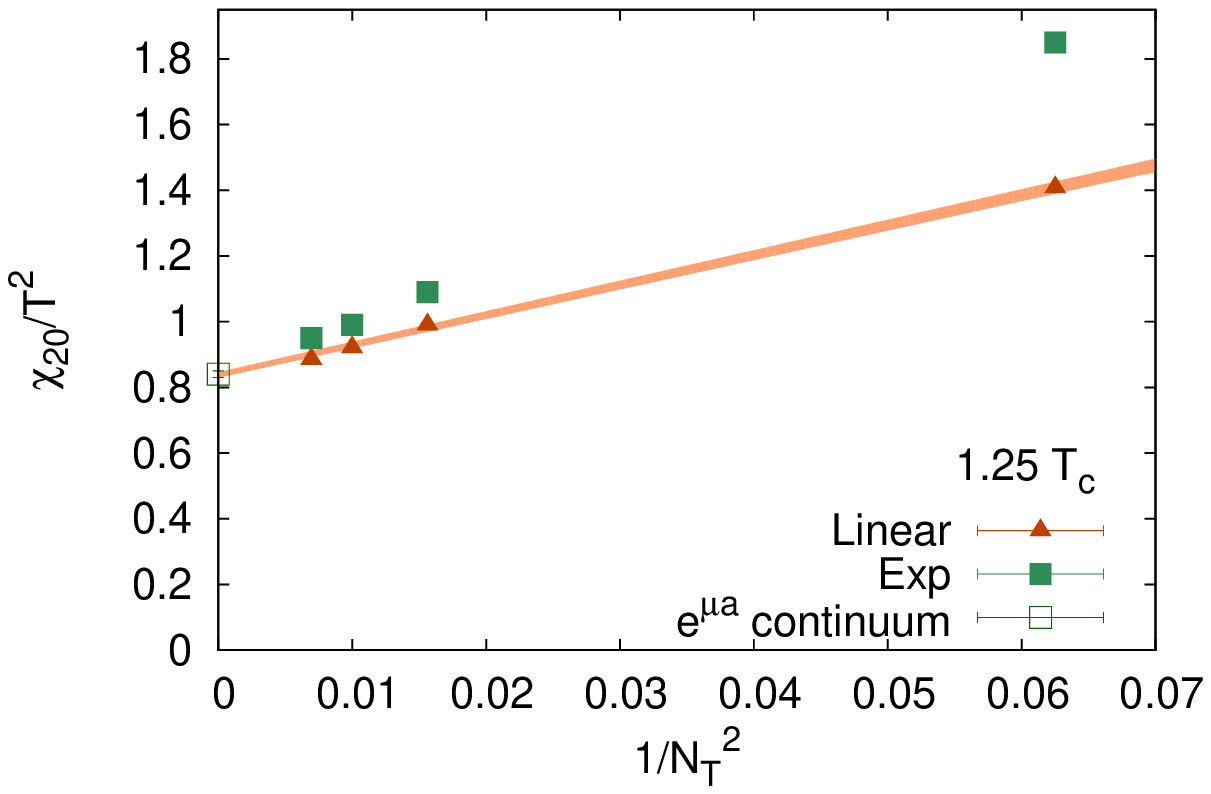}
\includegraphics[scale=0.52]{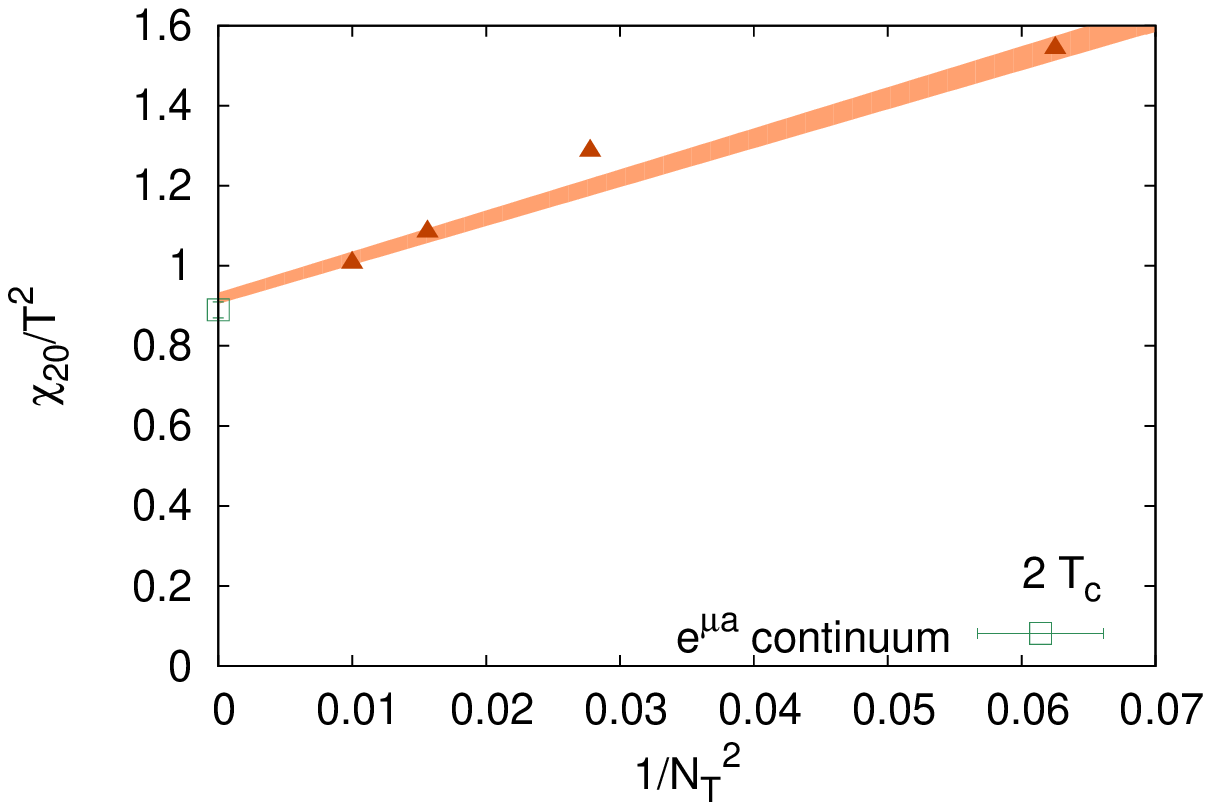}
\caption{The second order susceptibility at $1.25T_c$(left panel) and $2T_c$(right panel) for $m/T_c=0.1$. For 
comparing the cut-off effects we also include the available data from ~\cite{swagato} for the same quantity,  
calculated with the conventional exponential method. It is shown by squares in the left panel of the figure.}
\label{susc2} 
\end{center}
\end{figure}

\begin{figure}[h]
\begin{center}
\includegraphics[scale=0.52]{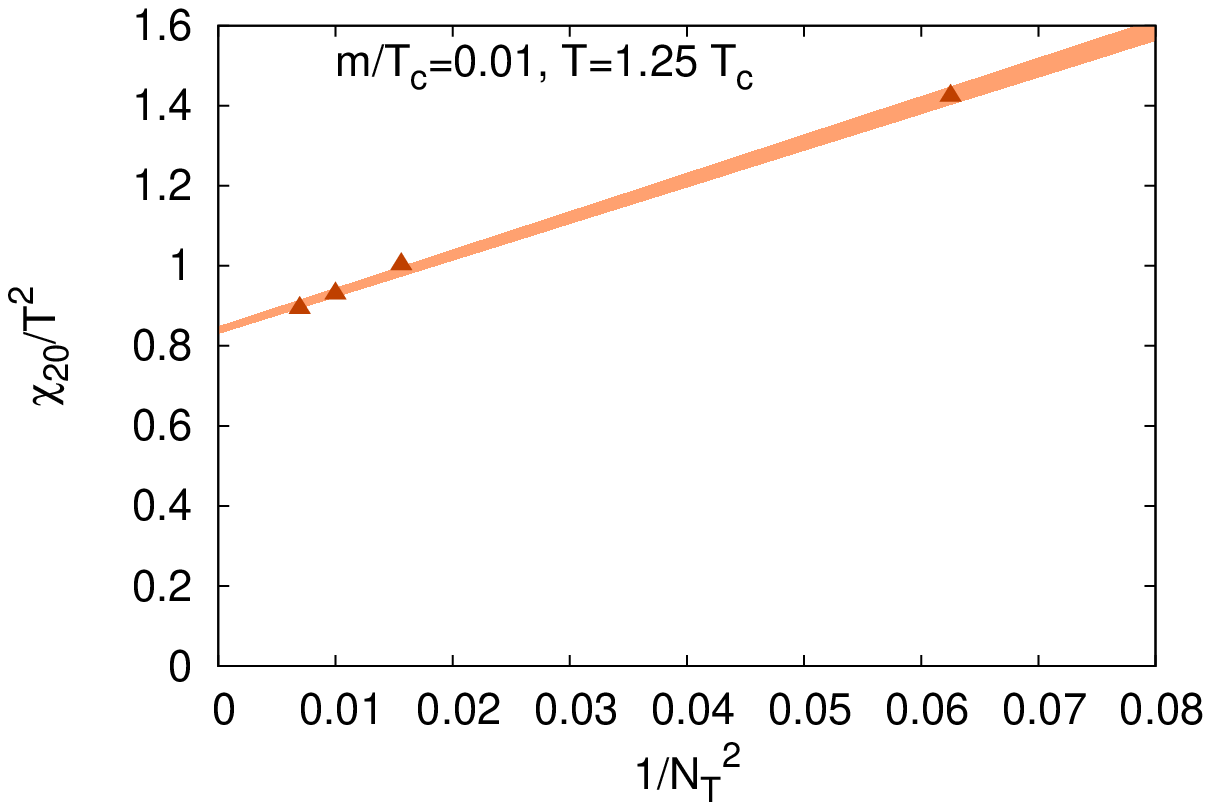}
\includegraphics[scale=0.52]{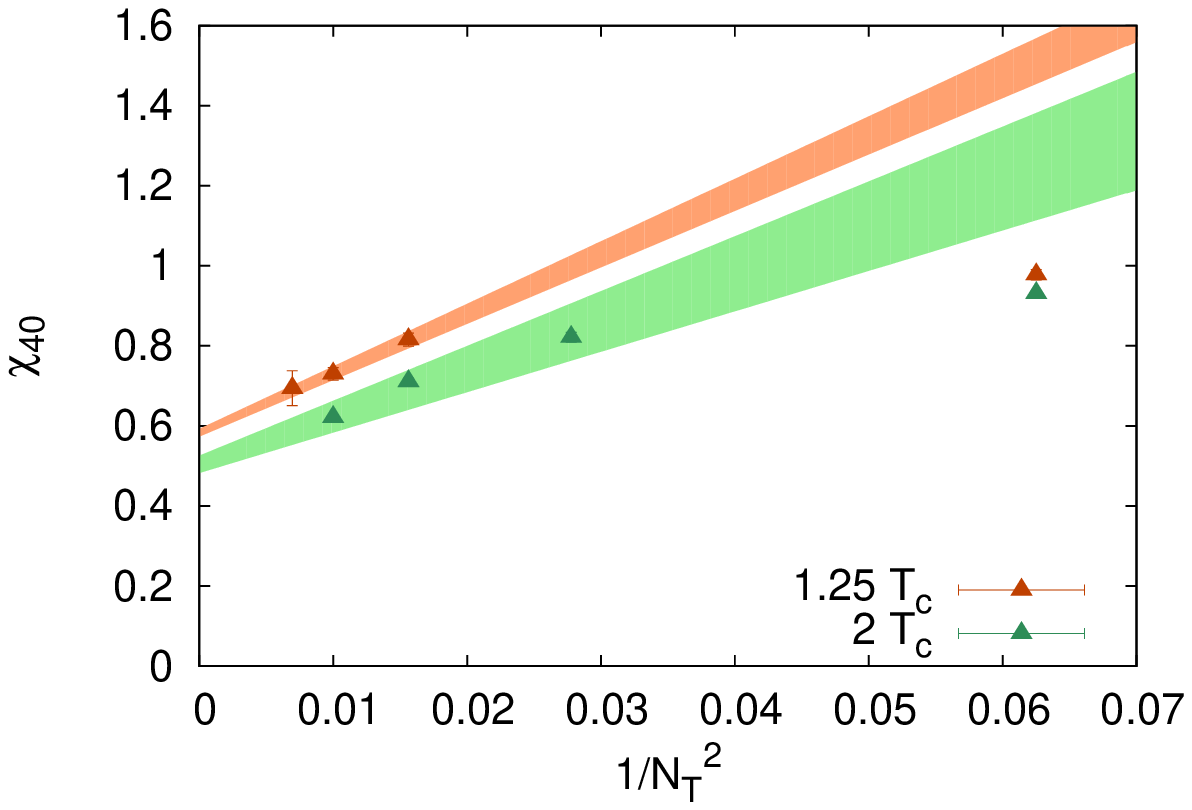}
\caption{The second order susceptibility at $1.25T_c$ shows no divergence in our method even for $m/T_c=0.01$ (left panel). Right panel 
displays the continuum extrapolated results for the diagonal fourth order susceptibility for $m/T_c=0.1$ at two 
different temperatures.}
\label{susc24d}
\end{center}
\end{figure}

The results for the dimensionless second order susceptibility $\chi_{20}/T^2$,
after the subtraction of the free results at $1.25~T_c$ and $2~T_c$
are displayed in Figure \ref{susc2} for the same physical quark mass $m/T_c=0.1$, 
and in the left panel of Figure \ref{susc24d} for the smaller $m/T_c =0.01$.  
For a comparison, we also plotted the available data from ~\cite{swagato}
of the same quantity at $1.25 T_c$ calculated using the conventional exponential 
method, in the left panel of Figure \ref{susc2}. The cut-off effects in the 
exponential method are larger than the linear method. The difference 
reduces on finer lattices, thus converging to the same continuum limit, in 
agreement with expectations from universality.
On general grounds, we expect $\chi_{20}/T^2$ should behave as
\begin{equation}
 \frac{\chi_{20}}{T^2}=c_1(T)+c_2(T) N_T^2+\frac{c_3(T)}{N_T^2}+\mathcal{O}(1/N_T^4)
\end{equation}

Since the values of the second order susceptibilities reduce with increasing 
$N_T$ in all the plots, there is clearly no sign of any divergence in
the interacting theory with $c_2(T) = 0$.  This is so irrespective of the
temperature, and even after lowering the quark mass by a factor of 10.  Our
results for the least squares fit of the data for the coefficients $c_1$ 
and $c_3$ are summarized in Table \ref{tab:table2}. For a proper comparison 
with the earlier data ~\cite{swagato}, we also included the 
$N_T=4$ points at both the temperatures. A measure of how good the least squares 
fit represents the data is given by a quantity $R^2$. It is defined as 
the ratio of the sum of squares in the fit model to the total sum of squares. 
The value of $R^2=1$ would therefore represent the best fit.
The values for our fit at $1.25 T_c$ and $2 T_c$ for $m/T_c=0.1$ are 
$R^2=0.9982$ and $R^2=0.9625$ respectively and that for $m/T_c =0.01$ is 
$R^2=0.9977$.
We also tried fits with a logarithmic term in Eq. (12) of the form $\ln(1/N_T^2)$ 
both in place of, and in addition to, the $1/N_T^2$ term. While the former seems 
strongly unfavoured, with an increase in $\chi^2$ by a factor of three to five, 
the latter is only marginally ruled out.  Additional lattices with larger $N_T$ 
are needed to be definitive for the latter case.  Thus either only a power law divergence or only 
 a logarithmic divergence is ruled out from our present data.
\begin{table}
 \centering
 \begin{tabular}{|c|c|c|r|r|}
 \hline \hline
 Observable& $m/T_c$ & $T/T_c$& $c_1$ &$c_3$\\  
 \hline
 & & 1.25& 0.838(8) & 9.1(2) \\  
 &0.1 & & &  \\
$\chi_{20}/T^2$ & &2 & 0.94(5)& 10(1)\\  
\hline
 &0.01 & 1.25 & 0.839(8) &9.4(2)\\  
 \hline \hline
 & & 1.25& 0.58(1) & 14.9(8)  \\ 
 $\chi_{40}$&0.1 & & &   \\
 & &2 & 0.50(2)& 12(2)   \\ 
\hline
 \end{tabular}
\caption{The continuum extrapolation results for the second and fourth order diagonal susceptibility.}
\label{tab:table2}
\end{table}

As seen in Figure \ref{susc2}, our continuum extrapolated values are in
agreement with the corresponding results obtained using the exponential in $\mu$
action \cite{swagato} at these temperatures. We also verified that there is no
mass dependent divergent term in the expression for the second order
susceptibility by performing a similar continuum extrapolation using a different
bare quark mass of $m/T_c=0.01$, as shown in left panel of Figure \ref{susc24d}. The second
order off-diagonal susceptibility is identical in both the linear and the
exponential method. Since it is zero for free fermion, no additional subtraction
is necessary for the linear case.

Calculations for free fermions show that the fourth order quark number
susceptibility has no divergent contributions but an additional finite
contribution in the continuum limit for the linear $\mu$-case.  Adopting the
same procedure for it as well, we subtract the additional  obtained free
contribution from the corresponding (quenched) lattice QCD determination.
For the fourth order susceptibility, there are one diagonal and two off-diagonal
components. Following the convention of \cite{gg1}, these can be written as,
\begin{eqnarray}
 \label{eqn:chi4}
 \nonumber
\chi_{40}&=&\frac{T}{V}\left[\langle\mathcal{O}_{1111}+6\mathcal{O}_{112}+4\mathcal{O}_{13}
+3\mathcal{O}_{22}+\mathcal{O}_{4}\rangle-3\langle \mathcal{O}_{11}+\mathcal{O}_{2}\rangle^2
\right]~,~\\\nonumber
\chi_{22}&=&\frac{T}{V}\left[\langle\mathcal{O}_{1111}+2\mathcal{O}_{112}+\mathcal{O}_{22}\rangle-
\langle \mathcal{O}_{11}+\mathcal{O}_{2}\rangle^2-2\langle \mathcal{O}_{11}\rangle^2\right]~,~\\
\chi_{31}&=&\frac{T}{V}\left[\langle\mathcal{O}_{1111}+3\mathcal{O}_{112}+\mathcal{O}_{13}\rangle-3
\langle \mathcal{O}_{11}+\mathcal{O}_{2}\rangle \langle \mathcal{O}_{11}\rangle\right]~,
\end{eqnarray}

where the operators $\mathcal{O}_n$ satisfy the identities $\mathcal{O}'_n =
\mathcal{O}_{n+1}$ and $\mathcal{O}_{ij} = \mathcal{O}_i\cdot \mathcal{O}_j$ and
so on.  The number density is given by $n = T \langle \mathcal{O}_1 \rangle/ V$.
$\mathcal{O}_2  = tr(-D^{-1}D'D^{-1}D') $  was the source of the divergence in
$\chi_{20}$, which  was cured by employing $\mathcal{O}_2-\mathcal{O}_2^{\rm
free~div}$, as discussed above. In order to be consistent, a subtraction of such
a constant from $\mathcal{O}_2$ in the expressions above should also be done.
It can be easily verified that such a substitution in the expressions above does
not change them at all since all the free theory divergence terms arising out of
it cancel in each expression.  This is, of course, consistent with the fact the
direct computation of the free diagonal susceptibility $\chi_{40}$ has no
divergence in the continuum limit for the linear $\mu$-case either. It does
have a constant $a^0$ term as an artifact though in the term coming from
$\mathcal{O}_4$.  Indeed, it is clear that due to dimensional reasons any
difference between the linear and exponential case must be a constant for
$\mathcal{O}_4$.  Moreover, all $\mathcal{O}_n$ for $n > 4$ will only differ by
terms which vanish in the continuum limit, being of order $a^{n-4}$.  Thus the
still higher susceptibilities must all agree in the continuum limit with the
exponential $\mu$-case.  Inspired by the success for the second order
susceptibility, we computed the $\mathcal{O}_4$ for free fermions at the same
temperature and subtracted it from the value obtained for the quenched theory at
that temperature in order to eliminate the $a^0$ term artifact.  Our results for
different $N_T$ are displayed in the right panel of Figure \ref{susc24d}.  These 
results for $\chi_{40}$ also show a convergence with increasing $N_T$. Thus there 
are indeed no additional divergences in the continuum limit, as anticipated. 
Moreover, the subtraction of the unphysical artifact appears to have been done 
correctly on each lattice size.  The continuum results for the exponential case 
are not available in the literature to facilitate a comparison unlike Figure 
\ref{susc2}. The results do show a converging trend though. The results for 
the continuum extrapolation are tabulated in Table ~\ref{tab:table2}. In all 
these fits the $N_T=4$ data has not been considered since it clearly stands out 
of the trend. The $R^2$ for these fits are $0.996$ and $0.978$
at $1.25 T_c$ and $2 T_c$ respectively.

\begin{figure}
\begin{center}
\includegraphics[scale=0.52]{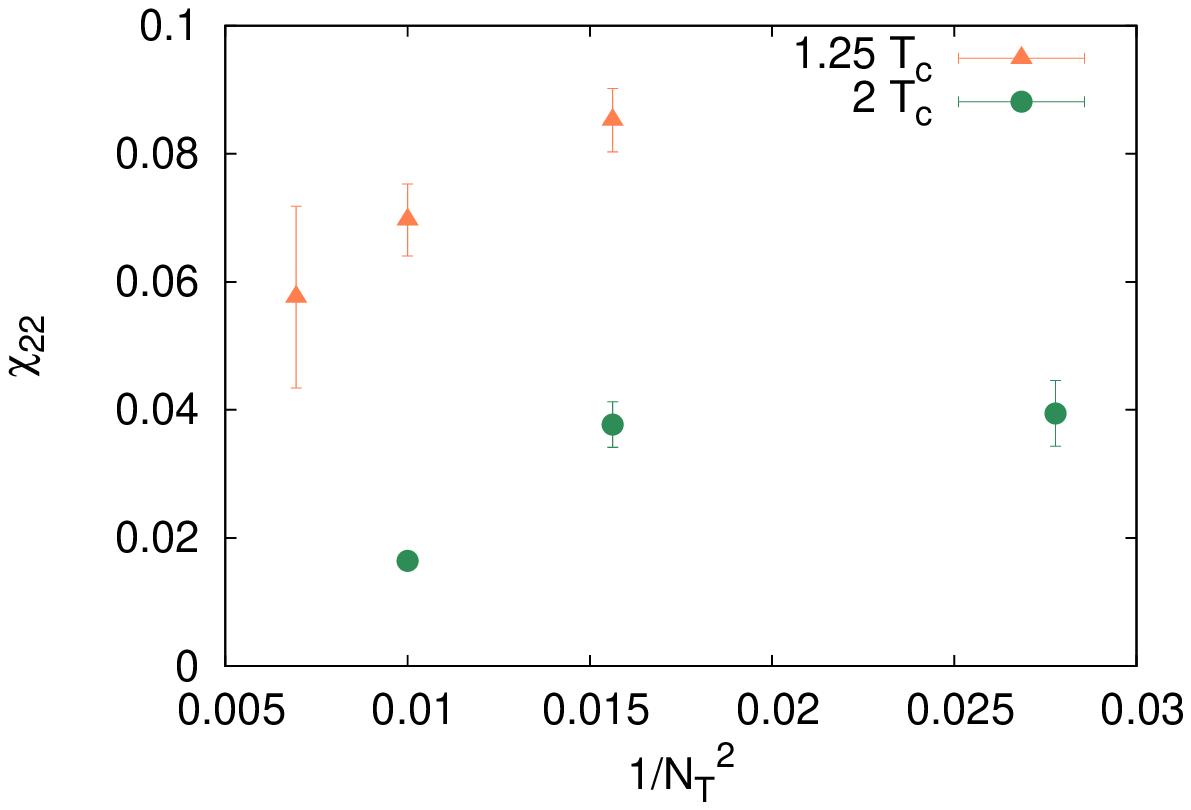}
\includegraphics[scale=0.52]{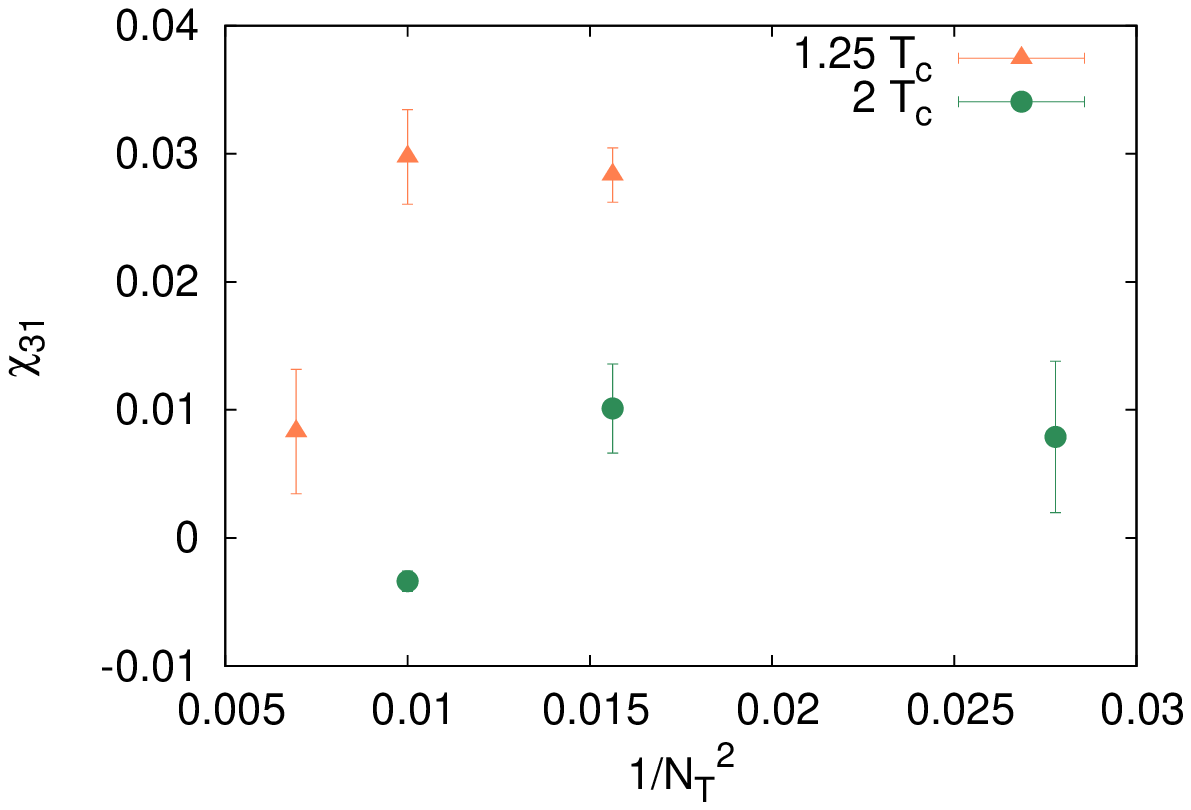}
\caption{The fourth order off-diagonal susceptibility $\chi_{22}$(left panel) 
and $\chi_{31}$(right panel) for $m/T_c=0.1$.}
\label{susc4offd}
\end{center}
\end{figure}

For the off-diagonal susceptibilities at fourth order, the free theory artifacts
are again zero so no subtraction are expected. Indeed, we observe this to be
true for $\chi_{22}$ in Figure \ref{susc4offd}. For $\chi_{31}$ in the right panel
of the same figure, it appears somewhat difficult to draw any definite conclusions although a 
finite continuum limit is suggested.

\section{Summary}
Investigations of QCD at finite density, and in particular of the QCD critical
point, gain from using the canonical Lagrange multiplier type linear chemical
potential term in the fermion actions on the lattice.   Preservation of exact
chiral invariance on the lattice seems feasible \cite{sharma} only for such a
linear term for the overlap and the domain wall fermions.  The higher order
quark number susceptibilities needed for locating the critical point using the
Taylor expansion approach are easier to compute in the linear case as well.
However, it is known \cite{hk,kogut,bilgav,gavai} since long that the linear
term  leads to  $\mathcal{O}(1/a^2)$ divergences in the baryon number
susceptibility.  We have shown that such a divergence exists already in the
continuum for a gas of free fermions, and therefore, lattice merely faithfully
reproduces it.  Using simulations of the quenched QCD with the staggered
fermions, we have verified that once the free theory divergence is explicitly
subtracted out, the susceptibility has no additional divergence in the continuum
limit.  This is only to be expected since the conserved charge, or number
density, does not get renormalized in an interacting theory.  Furthermore, its
extrapolated value in the continuum agrees very well at two different temperatures with the similar 
continuum results for the fermion action using terms
exponential in $\mu$, which by construction is free of any such divergences.
The higher order susceptibilities were also shown to be free of divergences in
quenched QCD, and it was argued why this was to be expected.  
Further work is clearly needed to check that this conclusion hold for the
theory with dynamical quarks as well, although one expects the maximum
possible difference to arise in the importance of the logarithmic term
whose numerical significance may not be felt even with the current best
precision.

An important consequence of our study is that this would enable computation of the higher order QNS in a 
significantly shorter time and with better control of errors, thereby enabling  use of ratios of still higher 
order QNS in locating the QCD critical point and a more precise equation of state at finite baryon density. 
These could also be exciting for the heavy ion experiments which have already
reported preliminary hints of a possible critical point and are beginning to
probe the finite density region in the ongoing and future programs.

\section{Acknowledgements}

This work was completed during the visit of one of us (RVG) to the
Universit\"at Bielefeld, Germany. He is very happy to acknowledge the kind
hospitality of its Physics Department, in particular that of Profs. 
Frithjof Karsch and Edwin Laermann.   R.V.G. also gratefully acknowledges the 
financial support from  the J.C. Bose National Fellowship of Department of 
Science \& Technology, Government of India (grant No. SR/S2/JCB-12/2009). 
S.S. would like to thank Robert Pisarski, Marc Sangel and S\"oren Schlichting for many 
helpful discussions.

\end{document}